\documentclass[prl,twocolumn,showpacs,preprintnumbers,amsmath,amssymb,superscriptaddress,floatfix]{revtex4}
\usepackage{graphicx}
\usepackage{dcolumn}
\usepackage{bm}
\begin{document}
\title{Mass measurements beyond the major \emph{r}-process waiting point $^{80}$Zn}
\author{S.~Baruah}
\affiliation{Ernst-Moritz-Arndt-Universit\"at, Institut f\"ur Physik, 17487 Greifswald, Germany}
\author{G.~Audi}
\affiliation{CSNSM-IN2P3-CNRS, Universit\'e de Paris Sud, Orsay, France}
\author{K.~Blaum}
\altaffiliation[Present address: ]{Max-Planck-Institut f\"ur Kernphysik, Saupfercheckweg 1, 69117 Heidelberg, Germany}
\affiliation{GSI, Planckstr. 1, 64291 Darmstadt, Germany}
\affiliation{Johannes Gutenberg-Universit\"at, Institut f\"ur Physik, 55099 Mainz, Germany}
\author{M.~Dworschak}
\affiliation{GSI, Planckstr. 1, 64291 Darmstadt, Germany}
\author{S.~George}
\altaffiliation[Present address: ]{Max-Planck-Institut f\"ur Kernphysik, Saupfercheckweg 1, 69117 Heidelberg, Germany}
\affiliation{GSI, Planckstr. 1, 64291 Darmstadt, Germany}
\affiliation{Johannes Gutenberg-Universit\"at, Institut f\"ur Physik, 55099 Mainz, Germany}
\author{C.~Gu\'enaut}
\affiliation{IN2P3-CNRS-CSNSM, Orsay Campus, France}
\author{U.~Hager}
\altaffiliation[Present address: ]{TRIUMF, 4004 Wesbrook Mall, Vancouver, British Columbia, V6T 2A3, Canada}
\affiliation{University of Jyv\"askyl\"a, Department of Physics, P.O. Box 35 (YFL), 40014 Jyv\"askyl\"a, Finland}
\author{F.~Herfurth}
\affiliation{GSI, Planckstr. 1, 64291 Darmstadt, Germany}
\author{A.~Herlert}
\altaffiliation[Present address: ]{CERN, Physics Department, 1211 Geneva 23, Switzerland}
\affiliation{Ernst-Moritz-Arndt-Universit\"at, Institut f\"ur Physik, 17487 Greifswald, Germany}
\author{A.~ Kellerbauer}
\altaffiliation[Present address: ]{Max-Planck-Institut f\"ur Kernphysik, Saupfercheckweg 1, 69117 Heidelberg, Germany}
\affiliation{CERN, Physics Department, 1211 Geneva 23, Switzerland}
\author{H.-J.~Kluge}
\affiliation{GSI, Planckstr. 1, 64291 Darmstadt, Germany}
\affiliation{Ruprecht-Karls-Universit\"at, Fakult\"at f\"ur Physik und Astronomie, 69120 Heidelberg, Germany}
\author{D.~Lunney}
\affiliation{CSNSM-IN2P3-CNRS, Universit\'e de Paris Sud, Orsay, France}
\author{H.~Schatz}
\affiliation{NSCL, Michigan State University, East Lansing, MI 48824-1321, USA}
\author{L.~Schweikhard}
\affiliation{Ernst-Moritz-Arndt-Universit\"at, Institut f\"ur Physik, 17487 Greifswald, Germany}
\author{C.~Yazidjian}
\affiliation{GSI, Planckstr. 1, 64291 Darmstadt, Germany}
\date{\today}
\begin{abstract}
High-precision mass measurements on neutron-rich zinc isotopes $^{71m,72-81}$Zn have been performed with the
Penning trap mass spectrometer ISOLTRAP. For the first time the mass of $^{81}$Zn has been experimentally
determined. This makes $^{80}$Zn the first of the few major waiting points along the path of the 
astrophysical rapid neutron capture process where neutron separation 
energy and neutron capture \emph{Q}-value are determined experimentally. As a consequence, the astrophysical conditions required
for this waiting point and its associated abundance signatures to occur in \emph{r}-process models
can now be mapped precisely. The measurements also confirm the robustness of the $N=50$ shell closure 
for $Z=30$ farther from stability.
\end{abstract}
\maketitle

The rapid neutron-capture process (\emph{r} process) is responsible for the synthesis of about half
of the heavy elements beyond Ge in the cosmos \cite{CTT91,TCC02,Qian03,ArnPRep}. However, where this process occurs is not known
with certainty. The characteristic abundance patterns of various competing models not only depend
on the chosen astrophysical environment, but also sensitively on the underlying nuclear physics processes and
parameters \cite{KBT93,CDK95,Wanajo04}.
Reliable nuclear data on the extremely neutron-rich nuclei participating in the \emph{r} process are
therefore needed to compare the signatures of specific models with
astronomical data now emerging from observations of metal-poor stars \cite{TCC02}.
Reliable nuclear-physics is also needed to disentangle contributions from many different processes
to the observed stellar abundance patterns below $A=130$, which were previously attributed to the \emph{r} process \cite{TGA04,MBC07}. 

$\beta$-decay half-lives and nuclear-mass differences along isotopic chains, i.e., neutron-separation energies, are of particular
importance in \emph{r}-process models, especially at the neutron-shell closures where the \emph{r} process passes through the $\beta$
decay of some particularly long-lived ``waiting-point" nuclei. The most prominent waiting points are believed to be
$^{80}$Zn, $^{130}$Cd, and probably $^{195}$Tm, responsible for producing the pronounced abundance peaks observed
around mass numbers  80, 130, and 195, respectively.  They serve as critical normalization points to constrain the conditions
required to produce a successful \emph{r} process \cite{KBT93}.

$Q$-value measurements related to these waiting points were performed only recently for $^{130}$Cd \cite{DKW03}
using the $\beta$-endpoint technique
and yielding a $\beta$-decay $Q$-value with an uncertainty of 150~keV. While these results provide important new
insights in nuclear structure and global mass models, their accuracy is not sufficient for \emph{r}-process calculations.
Furthermore, they
may have large systematic uncertainties due to poorly-known level schemes,
which can result in rapidly-increasing uncertainties
as measurements are extended to more exotic nuclei.
In addition, to constrain the path of the \emph{r} process, neutron-separation
energies and neutron capture $Q$-values are needed, requiring mass measurements beyond the waiting point nuclei with
a precision of the order of 10\,keV.

\begin{figure*}[t]
\begin{center}
\includegraphics[scale=0.8]{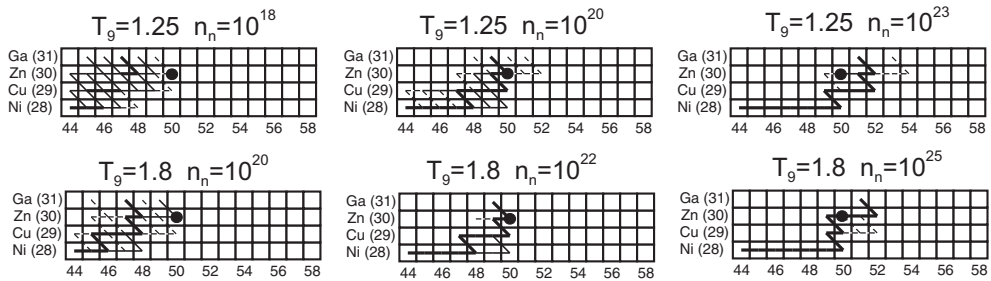}
\end{center}
\caption{\label{fig:1}
Reaction flows in the $^{80}$Zn (black dot) region calculated in this work for
various constant temperature and neutron-density conditions using the new nuclear
masses obtained in this work. Thick lines denote reaction flows (time integrated
abundance changes $dY=dX/A$ with mass fraction $X$ and mass number $A$) in excess
of $5 \times 10^{-3}$, while thin and dashed lines indicate reaction flows one or
two orders of magnitude lower, respectively.
}
\end{figure*}

In this letter, we report important experimental input for \emph{r}-process nuclear physics:  precision Penning-trap
mass measurements reaching the \emph{r}-process path at the $^{80}$Zn major waiting point,
including the first mass measurement of $^{81}$Zn.
Penning trap mass spectrometry has been developed in recent years for the use with radioactive beams,
and continued refinements have enabled mass
measurements with uncertainties of a few keV for increasingly exotic nuclei \cite{BlaumPhysRep,IJMSKlugeIssue}, which is
important for astrophysics calculations.
For example, Fig.\,\ref{fig:1} shows possible \emph{r}-process reaction flows in the $^{80}$Zn region.
Depending on the stellar conditions, 
the \emph{r}-process path either includes the slow $\beta$-decay of $^{80}$Zn making it a waiting point, or proceeds rapidly via neutron capture to 
$^{81}$Zn and beyond. For some more extreme conditions $^{80}$Zn can also be bypassed altogether.  
With the present measurements, $^{80}$Zn is the first major \emph{r}-process waiting point where experimental values of
the mass differences with respect to its neighbors ($^{79}$Zn and $^{81}$Zn) are available and thus a precise neutron-separation energy
as well as a precise neutron capture $Q$-value. The conditions
and role of $^{80}$Zn as a waiting point are discussed below.

The idea of a reduction of the strength of the closed neutron shell far from stability, discussed in the
literature as ``shell quenching'', has been evoked in the past in an attempt to explain some
\emph{r}-process abundance anomalies. However, evidence for this phenomenon in mass
regions relevant for the \emph{r} process has not been conclusive \cite{DKW03,Dworshak,Jungclaus}. 
The very recent mass measurements of Hakala et al. \cite{HakalaPRL} of neutron rich Zn masses 
up to $^{80}$Zn at JYFLTRAP indicate that the $N=50$ shell strength is somewhat diminished
down to $Z=31$. Here, we provide some evidence that the $N=50$ shell closure is retained
for the even more exotic case of $Z=30$.

The masses of $^{71m,72-81}$Zn have been determined with the Penning trap mass
spectrometer ISOLTRAP \cite{MukherjeeISOLTRAP} at ISOLDE/CERN. The radionuclides
were produced by impinging a 1.4-GeV proton pulse from the CERN Proton Synchroton
Booster with $3\times10^{13}$ protons every 2.4\,s on a uranium-carbide target.
In order to suppress a contamination by isobaric Rb isotopes, several
techniques have been used: First, the
protons were focussed on a neutron converter, i.e., a tungsten rod mounted next to
the target container, thus enhancing neutron-induced fission and reducing direct
nuclear reactions of protons with the actinide material.
Second, further suppression
of the alkali isotopes was achieved by introducing a quartz tube into the transfer line
between the target container and the ion source \cite{Bouquerel}: Rb isotopes diffusing
out of the heated target container freeze out on the surface of the quartz, while
the ions of interest pass unimpeded to the ionization region.
Third, the atoms were selectively ionized by
resonant laser ionization \cite{RILIS}.
In total, a suppression factor for $^{80}$Rb of the order of
$10^4$ was achieved at a quartz temperature of 680$^\circ$C \cite{Bouquerel}.

\begin{table}[b]
\caption{\label{tab:1}Experimental frequency ratios $r=\nu_{c,ref}/\nu_c$ of Zn$^+$ ions relative to
$^{85}$Rb$^+$. The mass excess $ME=(m-A)\cdot m_u$ (with mass number $A$ and mass $m$ in atomic mass units) has been deduced from the
mass $m=r(m_{ref}-m_e)+m_e$ using the conversion factor $m_u=931\,494.009\,0(71)$\,keV, the mass of $^{85}$Rb
$m_{ref}=84.911\,789\,738(12)$\,u, and the mass of the electron $m_e=548\,579.911\,0(12)\times10^{-9}$\,u \cite{AME2003}.}
\begin{ruledtabular}
\begin{tabular}{lccc}
Nuclide & frequency ratio $r$       & $\delta m/m$     & mass excess      \\
        &                           & $\times10^{-8}$  & exp. (keV)       \\
\hline
$^{71m}$Zn &  0.835\,311\,546\,0(302)  & 3.6              & -67\,171.2(2.4)  \\
$^{72}$Zn  &  0.847\,076\,232\,7(268)  & 3.2              & -68\,145.4(2.1)  \\
$^{73}$Zn  &  0.858\,885\,502\,0(236)  & 2.7              & -65\,593.4(1.9)  \\
$^{74}$Zn  &  0.870\,660\,440\,8(315)  & 3.6              & -65\,756.7(2.5)  \\
$^{75}$Zn  &  0.882\,477\,875\,0(245)  & 2.8              & -62\,558.9(1.9)  \\
$^{76}$Zn  &  0.894\,258\,121\,1(232)  & 2.6              & -62\,302.5(1.8)  \\
$^{77}$Zn  &  0.906\,079\,544\,4(287)  & 3.2              & -58\,789.1(2.3)  \\
$^{78}$Zn  &  0.917\,873\,056\,2(350)  & 3.8              & -57\,483.4(2.8)  \\
$^{79}$Zn  &  0.929\,701\,243\,6(490)  & 5.3              & -53\,435.1(3.9)  \\
$^{80}$Zn  &  0.941\,500\,837\,5(355)  & 3.8              & -51\,648.3(2.8)  \\
$^{81}$Zn  &  0.953\,346\,729\,7(632)  & 6.6              & -46\,199.6(5.0)  \\
\end{tabular}
\end{ruledtabular}
\end{table}

The Zn ions were accelerated to 60-keV kinetic energy
and further purified
by the high resolution mass separator HRS before being transferred to the
ISOLTRAP setup. The ions were stopped in a helium-gas-filled linear
radiofrequency quadrupole (RFQ) ion trap.
The ion bunch was transferred to the first Penning trap
for the removal of any isobaric contamination still present \cite{BlaumJPhysB}.
Finally, the purified
bunch of zinc ions was sent to the precision Penning trap, where the cyclotron
frequency $\nu_c=qB/(2\pi m)$ is measured with the well-established time-of-flight
ion cyclotron-resonance technique \cite{Graeff}.

In addition to isobaric cleaning, \emph{isomeric} purification by resonant dipolar rf excitation
\cite{BlaumEurPhysLett,VanRoosb} was used
in the precision trap in order to perform measurements on clean ion ensembles.
For $^{77}$Zn an admixture of the excited $^{77m}$Zn ($\Delta E=772.4$\,keV) needed to be
removed \cite{HerlertCzechJPhys}
to measure the cyclotron frequency of the ground state $^{77}$Zn$^+$. In the case of $^{71}$Zn, the
isomer was much more abundant than the ground state and therefore
the cyclotron frequency of $^{71m}$Zn$^+$ has been measured after removing any remaining
$^{71}$Zn$^+$ ions.
The determination of the cyclotron frequencies and the frequency ratios $r=\nu_{c,ref}/\nu_c$
of the reference ion $^{85}$Rb$^+$ and the ion of interest with the respective
uncertainties follows the procedure described in \cite{KellerbauerEPJD}. All present
uncertainties for the frequency ratios are governed by the statistical uncertainties of the
$\nu_c$ measurements, which are of the order of a few times $10^{-8}$.
For $^{81}$Zn, with a half-life of only 0.3\,s, the mass has been measured
for the first time and a relative mass
uncertainty below $7\times10^{-8}$ has been achieved.
The recent mass values of $^{76-80}$Zn measured by JYFLTRAP \cite{HakalaPRL}
are in excellent agreement.

In Table \ref{tab:1} the resulting mass excess values are given.
Some of them deviate from
the AME2003 literature values \cite{AME2003}, especially $^{73}$Zn, by more than four standard deviations.
Therefore, the present frequency ratios have been taken as input values for a new mass evaluation.
A new estimate for the mass excess of $^{82}$Zn was derived from the systematic trends of
the mass surface in that region of the nuclide chart. With $ME=-42\,860(500)$\,keV it is
shifted by -400\,keV as compared to the former estimated literature value \cite{AME2003}.
\begin{figure}[t]
\begin{center}
\includegraphics[angle=-90,scale=0.3]{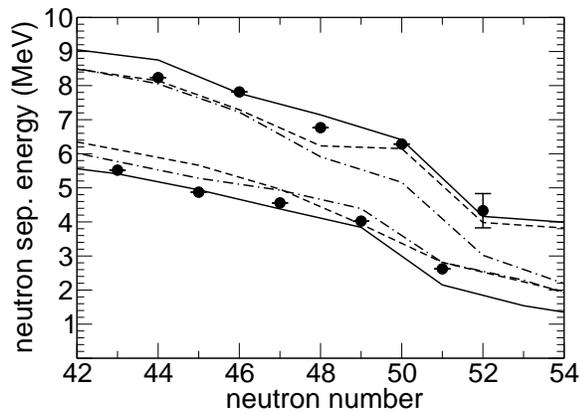}
\end{center}
\caption{\label{fig:3}
Neutron separation energies for neutron-rich Zn isotopes as functions of neutron number for the
FRDM (solid line) \cite{MNM95}, HFB-14 (dashed line) \cite{HFB14}, and ETFSI-Q (dot dashed line) \cite{ETFSIQmodel}
mass models together with the measured data from this work (black dots). }
\end{figure}
Figure~\ref{fig:3} shows the new neutron separation energies in the Zn isotopic chain as a function of
neutron number together with predictions from global mass models. The new data lead to accurate
neutron separation energies across the $N=50$ shell gap
and show that the strength of the $N=50$ shell closure is maintained for Zn.
This is in agreement with the recent study of the $N=50$ shell gap evolution \cite{HakalaPRL}
and extends its validity.
Amongst the mass models, the FRDM \cite{MNM95} shows the best
agreement with the data, and there is no indication of a reduction of the $N=50$ shell gap relative to this model.
The new HFB-14 mass model \cite{HFB14} also fits the data reasonably well. 
The ETFSI-Q mass model \cite{ETFSIQmodel} that had been developed to explore possible effects of shell
quenching clearly disagrees with the data.
\begin{figure}[b]
\begin{center}
\includegraphics[angle=-90,scale=0.3]{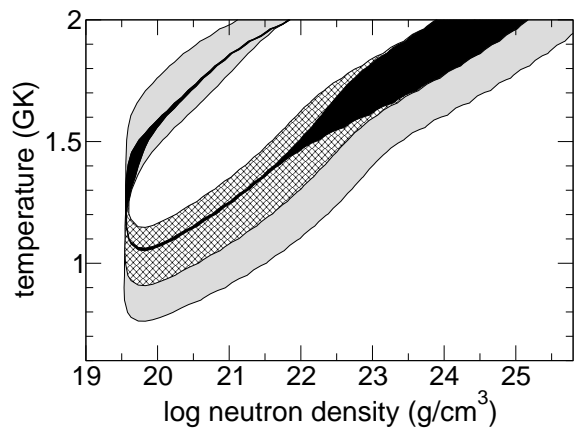}
\end{center}
\caption{\label{fig:4}
Boundaries enclosing the temperature and neutron density conditions required for an \emph{r} process to produce
a strong $A=80$ abundance peak. The black area indicates the result with mass values from this work. The
thickness of the boundary indicates the remaining mass related uncertainty. The cross hatched area indicates
the additional uncertainty without an improved mass value for $^{81}$Zn. The grey area indicates the additional
mass uncertainty when using the AME2003 mass values \cite{AME2003}.}
\end{figure}

For the subset of \emph{r}-process models that are characterized by a neutron-capture flow through
the $N=50$ mass region, the new results allow to determine the astrophysical conditions for which
$^{80}$Zn is a waiting point, therefore producing an $A=80$ abundance peak.
To that end we carried out dynamical reaction network calculations in the $^{80}$Zn region without the
assumption of (n,$\gamma$)$\rightleftharpoons$($\gamma$,n) equilibrium. The network includes isotopes from $N=44$ to 56
and elements from $Z=28$ to 31. (n,$\gamma$) reaction rates are taken from the statistical model NON-SMOKER \cite{RAU1}
using the FRDM mass
model \cite{MNM95}. The reverse ($\gamma$,n) reactions depend exponentially on neutron separation energies and
are calculated using the detailed balance principle (see \cite{CTT91} Eq. 4.5) with our new Zn mass values and
partition functions based on the same level densities and spin assignments as in NON-SMOKER.
With this approach we neglect the mass dependence of the (n,$\gamma$) reaction rates themselves.
This is typically not a strong dependence compared to the inherent uncertainty in the predicted rates. In addition,
for most of the relevant conditions, the \emph{r} process is dominated at least locally by (n,$\gamma$)$\rightleftharpoons$($\gamma$,n)
equilibria where the resulting reaction flows only depend on masses and partition functions, not the reaction rates themselves.
$\beta$-decay rates are taken from experiment when available \cite{AME2003,Hos05}, otherwise we use the
predictions from a global QRPA calculation \cite{MNK97}.

We start with an initial composition of $^{72}$Ni and neutrons only and run the network at a constant temperature and neutron
density until all the abundance is accumulated at $N=56$. Figure~\ref{fig:1} shows some of the results.
We then determine the fraction of the total reaction flow that has
passed through the $\beta$-decay of $^{80}$Zn for such conditions. To account for mass uncertainties,
we run for each temperature and neutron density multiple calculations for all possible permutations where
the Zn isotope masses are either set to their lower or their upper $1\sigma$ limit. From this we obtain the minimum and
the maximum fraction of flow through the $^{80}$Zn $\beta$-decay as a function of the temperature $T$ and the neutron density
$n_n$. Contours for 90\% $\beta$-decay branching are used to define the conditions under which $^{80}$Zn is a waiting point.

Figure~\ref{fig:4} shows the results for
AME2003 masses \cite{AME2003} and the new experimental data. Clearly the mass uncertainties
of 200-500~keV in the AME2003 extrapolations for the isotopes around $^{80}$Zn translate into
uncertainties of several orders of magnitude in neutron density and 0.2-0.4~GK in temperature. 
With our new data, in particular with the first precision mass value of $^{81}$Zn, the
picture changes drastically and we have now
a well defined map of conditions for a major \emph{r}-process waiting point to be on the reaction path.
This provides a first reliable guide for \emph{r}-process models with neutron capture in this mass region. Model
conditions need to pass through our defined region of temperature and neutron density in order to produce
the observed high $A=80$ abundance.

Figure~\ref{fig:4} shows two distinct temperature regimes above and below about 1.5~GK with different temperature 
and density dependencies and different sensitivities to mass uncertainties.  Above 1.5~GK 
(n,$\gamma$)$\rightleftharpoons$($\gamma$,n) equilibrium is fully established and even $N$ isotopes play the dominant
role because of their higher neutron separation energies. In this regime, the calculations are expected to be insensitive
to neutron capture rates. Additional uncertainties from partition functions should be small as well, as the even $N$ 
isotopes are even-even nuclei with ground state spin zero, and with a relatively low level density on excitation 
energy scales corresponding to the few 100 keV of thermal energy. The largest remaining uncertainty in this regime is the 
mass of $^{82}$Zn, which governs the leakage out of $^{80}$Zn. While our measurements 
have improved the mass extrapolation for $^{82}$Zn, a precision measurement would be desirable.
 
For temperatures below 1.5~GK  (n,$\gamma$)$\rightleftharpoons$($\gamma$,n) equilibrium begins to break down. For the 
high neutron density boundary, leakage out of $^{80}$Zn is governed by a local equilibrium between $^{80}$Zn and 
$^{81}$Zn, followed by neutron capture on $^{81}$Zn. The most critical quantity in this regime is the mass difference between 
$^{80}$Zn and $^{81}$Zn, which we have determined here experimentally for the first time making the previously dominating
mass uncertainties negligible.
For the low neutron density boundary
the reaction flow is determined by competition between $\beta$-decay and neutron capture and becomes insensitive to masses as
Fig.~\ref{fig:4} shows. For this low temperature regime, predictions of neutron capture rates 
and partition functions might contribute small additional uncertainties.

\begin{acknowledgments}
This work was supported by the German Federal Ministry for Education
and Research (06GF151 and 06MZ215),
by the European Commission (NIPNET RTD network HPRI-CT-2001-50034),
and by the Helmholtz association for research centres (VH-NG-037).
H.S. is supported by NSF grants PHY0606007 and PHY0216783.
S.B. thanks the University of Greifswald for a stipend in the framework
of an International Max Planck Research School.
We thank the ISOLDE Collaboration
and the technical team for their support.
\end{acknowledgments}

\end{document}